\documentclass[12pt]{iopart}
\usepackage{graphicx}
\usepackage{enumerate}

\usepackage{iopams}
\usepackage{epsfig}
\usepackage{subfigure}
\usepackage{hyperref}

\usepackage{color}
\usepackage[usenames,dvipsnames]{xcolor}

\newcommand{\RNum}[1]{\uppercase\expandafter{\romannumeral #1\relax}}

\begin{document}

\title{Differential Bloch Oscillating Transistor Pair}

\author{Jayanta Sarkar$^1$, Antti Puska$^1$, Juha Hassel$^2$ and Pertti J. Hakonen$^1$}
\address{$^1$Low Temperature Laboratory,  O.V.Lounasamaa Laboratory, \\Aalto University, P.O. Box 13500, FI-00076 AALTO, Finland}
\address{$^2$VTT Technical Research Centre of Finland, P.O. Box 1000, FI-02044 VTT, Finland}

\ead{jay@boojum.hut.fi}

\date{\today} 
%\maketitle

\begin{abstract}

We examine a Bloch Oscillating Transistor pair as a differential stage for cryogenic low-noise measurements. Using two oppositely biased, nearly symmetric Bloch Oscillating Transistors, we measured the sum and difference signals in the current gain and transconductance modes while changing the common mode signal, either voltage or current. From the common mode rejection ratio we find values $\sim 20$ dB even under non-optimal conditions. We also characterize the noise properties and obtain excellent noise performance for measurements having source impedances in the M$\Omega$ range.

\end{abstract}
%\pacs{1315, 9440T} 
%\submitto{\NT}
\maketitle
 
\section{Introduction}
Bloch Oscillating Transistor (BOT) was initially conceived as a quasiparticle to N Cooper pairs converter in which the current gain $\beta$ is given by $2N+1$ \cite{science}. 
However, it was soon realized that incoherent tunneling plays a crucial role and coherent Bloch oscillations were impossible to reach without a resistive environment which was unrealistic from technological point of view \cite{LZ,devoret90,IN,averin,lindellJLTP}. In regular BOT devices, substantial current gains ($\beta=\Delta I_{out}/\Delta I_{in}$ )
 have been hard to achieve \cite{science,lindellJLTP}. Recently, it has been demonstrated that large $\beta_E$-values can be obtained in devices whose characteristics are close to bistable behavior \cite{jayLONG}. The base current can have two different steady-state values in the hysteretic regime and, consequently, the onset of hysteresis corresponds to a bifurcation point where the current gain of the device diverges. 

 The applications of a BOT are expected to be mostly in metrology \cite{piquemal}, dealing with source resistances of a few M$\Omega$ at low frequencies below 1 kHz \cite{seppametrology,seppametrology2}.
One of the remarkable features of all BOTs is that they may operate below the shot noise limit as clearly demonstrated in Ref. \cite{lindellAPL} where a suppression factor of 4 was reached at base current 60 pA and input referred current noise, $i_n \sim 1$ fA/${\rm \sqrt{Hz}}$ was obtained. In this work we further suppress the shot noise by choosing to work with BOT devices in which the onset of hysteresis can be tuned at small base currents. When operated near the onset of hysteresis, the regular current amplifier description is insufficient with realistic source impedances as the input impedance and the optimum noise impedance are also expected to diverge along with current gain \cite{hassel04}. Hence, the relevant figures of merit are the transconductance gain and the voltage noise which are experimentally characterized in the present work.

One of the challenges of high-sensitivity, low-noise measurements is their susceptibility against external perturbations. As a consequence, differential amplifier stages are commonly preferred since they can eliminate common mode disturbances and detect only the differential signal from the source. In this paper, we also study the applicability of  BOTs for differential measurements. We have connected two BOTs in a manner where we have been able to bias them separately with opposite bias voltages (currents) and measure the sum and difference signals in current and transconductance gain modes while changing the common mode (CM) signal, either voltage or current. We characterize the results in terms of common mode rejection ratio (CMRR) which is found to be 20 dB even without properly optimizing the operating point in a slightly asymmetric device. Moreover, we compare the noise properties of the BOT in current and transconductance gain modes.

\section{Experimental details}
\subsection{Sample Fabrication}
The BOT samples employed in this work were fabricated on oxidized Si substrates. LOR 3B was spun at 2000 rpm for 60s and baked on hot plate at 150 C for 10 mins. We repeated it for two times which gives total thickness of LOR of $\sim$ 800 nm. 20 nm thick Ge was thermally evaporated at a rate of 0.3 \AA/s. PMMA was spun at 4000 rpm on Ge layer followed by baking for 3 mins at 170 C. Rate of deposition of Ge and baking of PMMA found to be two important factors which play a role to generate cracks in Ge film. 

The BOT structures were patterned on this layer using e-beam lithography at 20 keV. After pattering the chips were subjected to MIBK:IPA (1:3) solution for developing, followed by reactive ion etching with  ${\rm CHF_4}$ plasma to etch away the Ge through the PMMA window.
Finally, the LOR under the germanium was etched in oxygen plasma up to the desired extent of undercut in an inductively coupled plasma etcher with temperature control for the substrate holder.

Shadow angle evaporation at four different angles was employed to generate the structures consisting of three metals.
The process order in the evaporation sequence was (\RNum{1}) Chromium, (\RNum{2}) Aluminum, (\RNum{3}) oxidization, (\RNum{4}) Aluminum, and (\RNum{5}) Copper. Oxidation was done in Ar:O$_2$ (6:1) mixture at 80 mTorr for 1 min. NMP or PG remover was used for lift-off. 

In short, Al-${\rm AlO_{x}}$-Al SQUID forms the emitter electrode while a Cr thin film resistor works as the collector, and normal metal - insulator - superconductor (NIS) junction corresponds to the base. Even though the nominal process parameters for the BOTs on the same chip were equal, the characteristics of the individual devices varied substantially, as can be seen from Table \ref{BOTparams} which summarizes the parameters for the measured differential pair. The underlying cause is attributed to the nonuniform bending of the Ge mask after the etching that was employed to create the undercut. Due to this reason, angle evaporation of different metals did not give exactly identical sample structures among all the samples on the chip.

\begin{figure}[!h]
\center
\includegraphics[width = 9cm,height=8.05cm]{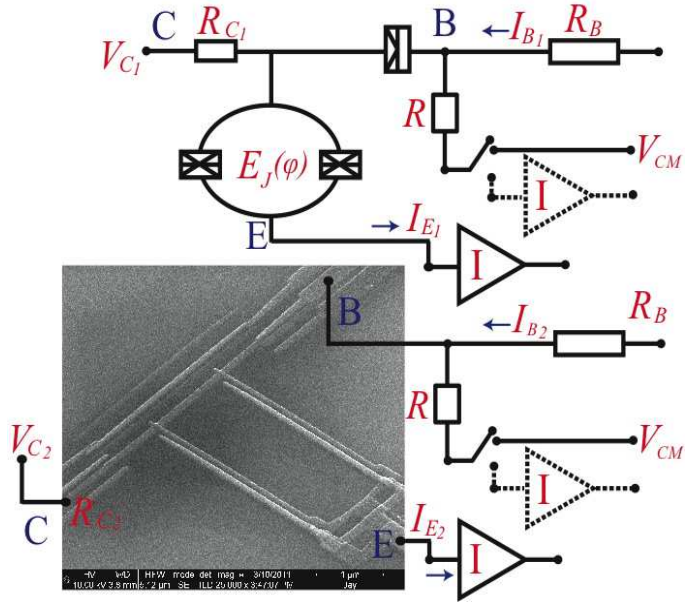}
\caption{Schematic layout of the setup used in the differential mode measurements as well as in the characterization of the individual BOTs. The integrated SEM image illustrates one of our BOT devices that were obtained using four angle evaporation. In both upper and lower branch pictures, base, emitter, and
collector are marked by B, E, and C, respectively. $V_{CM}$ port is used for applying the common mode signal to the BOTs with the switches in the upper position.
}\label{BOTschema}
\end{figure}

\begin{table}[!h]
\center
\begin{tabular}{lllllllll}
\hline\noalign{\smallskip}
BOT \#&$R_{N} (k\Omega)$&$R_{JJ}(k\Omega)$&$R_C(k\Omega)$& $E_J(\mu eV)$&$E_C(\mu eV)$\\
\noalign{\smallskip}\hline\noalign{\smallskip}
1 & 53 & 27  & 550 & 17 & 40  \\
2 & 123 & 19  & 660 & 18 & 45  \\
\noalign{\smallskip}\hline
\end{tabular}
 \caption{Parameters for the measured BOT samples. $R_{N}$ and $R_{JJ}$  are the normal
state resistances of the NIS and JJ tunnel junctions (SQUID),
respectively. $E_J$ denotes the Josephson energy of the SQUID and $E_C$ is the charging energy associated with the total capacitance ({\it C}) of the system.
 }
\label{BOTparams}
\end{table}

\subsection{Measurements}
The measurements were done on a plastic dilution refrigerator (PDR-50) from
Nanoway Ltd. The base temperature of the refrigerator was 50 mK. The filtering in the PDR
consisted of 70 cm long Thermocoax cables on the sample
holder and 1 k$\Omega$ series resistors at 1.5 K. Also, microwave filters from Mini-circuits (BLP 1.9) were
used at the top of the cryostat in the bias lines. The electronic temperature, deduced from the $IV$ characteristics,
turned out to be 150 mK, which is presumably due to Joule-heating of the extremely thin Cr resistors.

Aluminum bonding wires were used to connect the E, B and C (see Fig. \ref{BOTschema}) contact-pads to the sample holder. 
Surface-mount, metallic thin film resistors of $R = 500$ k$\Omega$ (510 k$\Omega$ at 100 mK) were soldered on the sample holder itself; the whole configuration was cooled to base temperature to minimize the thermal noise coming from the resistors $R$. Two $R$s were connected to common-mode port with thin copper wire.
 In the transconductance $g_m=-\frac{\Delta I_E}{\Delta V_B}$ and input impedance measurements $Z_{in}=\frac{\Delta V_B}{\Delta I_B}$, we measured the current through $R$, which allowed us to determine the current division of $I_{B1}$ and $I_{B2}$ at the base. Knowledge on the division allowed us to determine $Z_{in}$ and $g_m$ of the BOTs simultaneously.

The base electrodes of the BOTs were DC biased, either by current or voltage depending on the configuration on the CM port. Base currents were fed via equally
large resistors $ R_B = 1-10$ G$\Omega$, located at room temperature. Ground contact at the other end of $R$ converted the current bias into voltage bias. Voltages at the base and at the collector were tracked by  LI-75A
amplifiers while currents were measured using DL1211 preamplifiers. For noise measurement, SR780 spectrum analyzer was employed.

\section{Results and Discussion}
First, we characterized both samples in the current bias mode and measured the current gain ($\beta=-\frac{\Delta I_E}{\Delta I_B}$) on the $I_B$ vs. $E_J$ plane. Similar cross-over curve between stable and bistable operation was observed as found in Ref. \cite{jayLONG}. Near the bifurcation point, large gains were obtained and we chose to operate in a regime where the cross-over curve is rather insensitive to changes in $E_J$, which was expected to improve the long term stability of the CMRR measurement. Close to the divergence point, it is also easy to tune the current gains to $\beta_{1}=\beta_{2} \sim 20$ for the CMRR measurement. Unfortunately, when the CM port was fed using current, the CMRR  was found to be poor. This was traced back to the  difference in $R_C$ values of the BOTs, because an equally large relative difference is there in the input impedances as $Z_{in} \simeq \beta_E R_C$. The main obstacle in balancing between $Z_{in}$ values is the  variation in the $\sim$ 6.5 nm thick chromium resistor which is difficult to control within a few percent in our lithographic process. Therefore, the common mode current is divided unevenly between the BOTs and no proper cancellation is achieved. Hence, we conclude that the operation of a
differential BOT pair is more difficult in current gain mode than in
transconductance regime where it is sufficient to  match just the gains.  %\cite{obstacle}. 

Second, we grounded the resistors $R$ and traced out the transconductances of the two BOTs ($R_{B}=5$ G$\Omega$). In order to measure $g_m$ along the whole $I_E-V_C$ characteristics, we traced $I_E$ using different values for $I_B$. $\Delta  I_E$ was calculated by subtracting two $ I_E$ traces for two different base currents, whereas $ \Delta V_B$ was determined from the change in the division of $I_{B1}$ ($I_{B2}$).

\begin{figure}
\center
\includegraphics[width = 8cm, height=9cm]{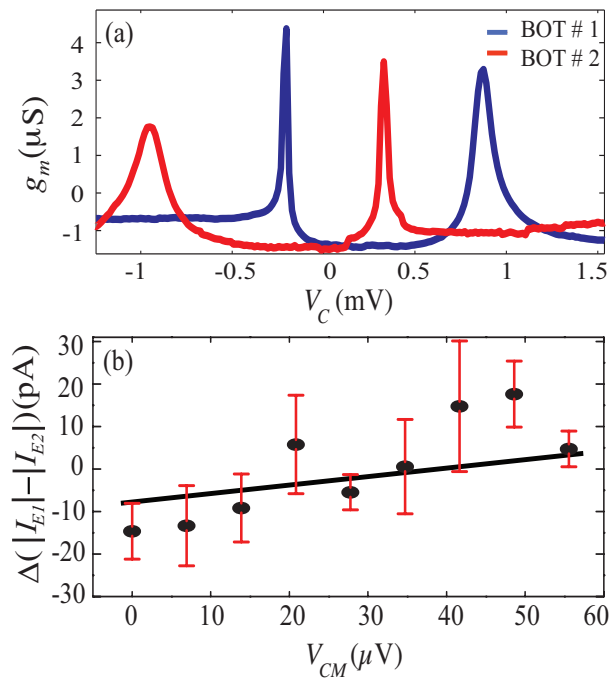}
\caption{a) Transconductance $g_m$ of the BOTs vs. collector bias $V_C$. At the operating point $g_m$ is at maximum.
b) Difference of absolute output currents of the BOTs, $\Delta(|I_{E1}|-|I_{E2}|)$ plotted vs. voltage applied to the CM port. The straight line yields a CMRR of 20 dB. 
}
\label{CMdata}
\end{figure}

 Fig. \ref{CMdata}a shows the measured $g_m$ for both BOTs plotted against voltage $V_C$. The absolute maximum value of transconductance peaks appear in the negative resistance regime of $I_E-V_C$ curves. The maximum for $g_m$ is found to vary from 2 $\mu$S to 9 $\mu$S in the present samples which corresponds to $\frac{1}{R_C}$ to $\frac{4}{R_C}$. This coincides well with the simulations of Ref. \cite{hasselLONG}. 
The transconductance was recorded across the hysteretic point of the current gain, but no abrupt change was found in $g_m$. This finding reflects the fact that both $Z_{in}$ and $\beta_E$ diverge and $g_m=\frac{\beta_E}{Z_{in}}$ \cite{hasselLONG}.

For a differential BOT we define the signal to be the emitter difference current of the BOTs. In a realistic configuration the post-amplification could be arranged, e.g., by transforming the currents to magnetic fluxes with opposite polarities in a SQUID postamplifier. Thus we can define the CMRR as $-20 {\rm log}\left (2\frac{|g_{m1}|-|g_{m2}|}{|g_{m1}|+|g_{m2}|}\right)$ when the magnitude of the base bias voltage is the same. Due to opposite bias in the non-hysteretic regime ($g_{m} > 0$) we can have $g_{m1} \simeq g_{m2}$ and $I_{E1} \simeq -I_{E2}$ ($V_{B1} \simeq -V_{B2}$). If $g_{m1} = g_{m2}$, $I_{E1} - I_{E2}$ would be fully independent of the common mode signal and CMRR = $\infty$.

Fig. \ref{CMdata}b shows one example of the effect of $ V_{CM}$ on the output current $I_{E1} - I_{E2}$. In this case, the two BOTs were biased at the points where they have almost equal magnitude of transconductance, $g_{m1}=1.9$ $\mu$S and $g_{m2}=2.1$ $\mu$S, but the values of which may increase slightly with $V_{CM}$. Nevertheless, as $V_{CM}$ is varied, $\Delta (|I_{E_1}|-|I_{E_2}|)$ changes rather weakly, and the straight line in Fig. \ref{CMdata}b yields a CMRR value of 20 dB.

\begin{figure}
\center
\includegraphics[width = 8.4cm,height=8cm]{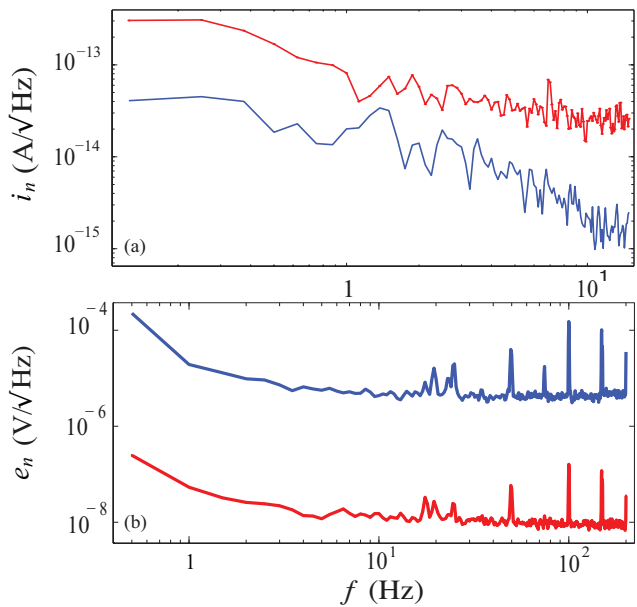}
\caption{a) Input referred current noise ($ i_n$) in the current gain mode: red trace displays $i_n$ with $\beta_E = 1$  i.e., far from the active operating regime. The blue curve depicts $i_n$ measured at $\beta_E = 35$. b) Input referred voltage noise ($e_n$) in the transconductance mode: the blue curve is obtained at $g_{m}= 10$ nS while the red trace was measured at $g_{m}= 10$ $\mu$S.
}\label{noiseintrans}
\end{figure}

We also performed equivalent input noise measurements on BOTs in the transconductance mode, connected to a 510 k$\Omega$ source impedance. Since the source impedance is clearly less than the impedance for noise optimum, we are mostly sensitive to the voltage noise generator of the amplifier. Fig. \ref{noiseintrans} shows the noise spectral density on the emitter terminal under current and voltage biased conditions. Fig. \ref{noiseintrans}a displays the input referred current noise when the device is used as a current amplifier. The blue trace is at the operating point with $\beta_E=35$. The bandwidth was found to be $\sim$ 20 Hz, 
which is consistent with the $RC$ input time constant with $Z_{in} \sim$ 10 M$\Omega$. The input referred current noise exceeds $i_n \sim {\rm 1fA}/{\rm \sqrt{Hz}}$ below 10 Hz, which indicates slightly more $1/f$ noise compared with Ref. 
\cite{lindellJLTP}.

 The  equivalent input voltage noise $e_n$ in the transconductance mode is displayed in Fig. \ref {noiseintrans}b. We find $e_n\sim$ 10 nV/${\rm \sqrt{Hz}}$ at the maximum  $g_m = 10$ $\mu$S. The bandwidth is around 200 Hz which is limited by the output impedance of $\sim 500$ k${\rm\Omega}$. Simultaneously with these noise measurements, we determined the input impedance and obtained $Z_{in}$ = 6 M$\Omega$. Using the obtained $e_n$ and $Z_{in}$, we estimate for the optimum impedance $Z_{opt}$=5 - 10 M$\Omega$. The uncertainty in $Z_{opt}$ arises from the different operating conditions in the determination of $e_n$ and $i_n$. In order to deduce $Z_{opt}$ we have made the assumption that the equivalent input noise of the BOT near the bistability point is governed by the output switching noise, which allows scaling of the noise sources between different conditions. This  estimate for optimum impedance supports the theoretical result $Z_{in}\approx Z_{opt}$  of Ref. \cite{hassel04}.

In the transconductance mode the output current noise is $g_m e_n \approx$ 100 fA/${\rm \sqrt{Hz}}$, which sets the criterion for the post-amplification. This requirement can be met, e.g., with SQUID-based post-amplifier \cite{SQ1,SQ2}. 
Within the source resistance range 1 M $\Omega$  
$<R_s \lesssim Z_{opt} $, the BOT transconductance amplifier over-performs SQUID-based amplifiers as $e_n/R_s$ is in the range of 1-10  fA/${\rm \sqrt{Hz}}$, i.e., lower than the current noise of the state-of-the-art SQUID ammeters at low frequencies. 

\section{Conclusion}
In conclusion, we have performed transconductance (current gain), voltage noise (current
noise), and  input impedance measurements in voltage (and current) biased,
galvanically coupled Bloch oscillating transistors  in  a differential pair
configuration. We found that biasing with voltage works better for the
CMRR measurement, providing easily a rejection
ratio of 20 dB. The superiority of the transconductance mode is attributed to the fact that it is then sufficient to match just the $g_m$-values of the
two amplifiers, while careful matching of $Z_{in}$ in addition to gain
$\beta_E$ is necessary in the current gain mode. In the transconductance mode,
we find for the input referred voltage noise $e_n = 10$ nV/${\rm \sqrt{Hz}}$
at $g_{m}=10$ $\mu$S and  $Z_{in}=6$ M$\Omega$, with the $1/f$ corner
frequency at around 5 Hz. Altogether, we can say that by matching the $g_m$ values of the two BOTs  closer to each other coupled BOTs will make  an excellent candidate for a transconductance amplifier in metrological cryogenic applications.
\section*{Acknowledgments}
We thank  Mikko Paalanen and Heikki Sepp\"a for fruitful discussion. We acknowledge Micronova cleanroom facilities for fabrication of our samples.
Financial support by Academy of Finland, Technology Industries of Finland Centennial Foundation and TEKES is gratefully acknowledged.

\end{document}